%% file: lat98.tex
\documentclass[twoside,fleqn]{article}
\usepackage{espcrc2}
\usepackage{epsfig}
\usepackage{latexsym}
\usepackage{epic}
\usepackage{eepic}

\title{
\vspace{-3.5cm}                         
{\normalsize DESY 98--147}     \\[-0.2cm]      
{\normalsize FU-HEP/98--7}    \\[-0.2cm]      
{\normalsize HUB--EP--98/62}   \\[-0.2cm]      
{\normalsize TPR--98--27}   \\[-0.2cm]      
{\normalsize HLRZ 1998--58}   \\[-0.2cm]      
{\normalsize September 1998}      \\              
\vspace{0.7cm} 
Towards a Non-Perturbative Calculation of DIS Wilson Coefficients%
\thanks{
Talk presented by D.~Petters at Lattice 1998.    
}}

\author{
S.~Capitani%
\address{Deutsches Elektronen-Synchrotron DESY,
                    D-22603 Hamburg},
M.~G\"ockeler%
           \address{Institut f\"ur Theoretische Physik, Universit\"at
                    Regensburg, D-93040 Regensburg},
        R.~Horsley%
           \address{Institut f\"ur Physik, Humboldt-Universit\"at zu Berlin,
                    D-10115 Berlin},
        H.~Oelrich%
           \address{Deutsches Elektronen-Synchrotron DESY and NIC, 
           D-15735 Zeuthen},
        D.~Petters$^{\rm d, }$%
           \address{Institut f\"ur Theoretische Physik,
                    Freie Universit\"at Berlin, D-14195 Berlin},
        P.~Rakow$^{\rm b}$, and
        G.~Schierholz$^{\rm a, d}$
}

\begin{document}

\begin{abstract}
We verify the operator product expansion (OPE) of deep inelastic scattering
(DIS) on the lattice and present first results of a non-perturbative 
calculation of the Wilson coefficients. 
\end{abstract}

\maketitle

\section{INTRODUCTION}
 
The calculation of power corrections in DIS requires not only the
evaluation of the matrix elements of higher-twist operators, but also
the computation of the Wilson coefficients beyond perturbation
theory~\cite{M&S}. In~\cite{renorm} we saw already that the scale 
dependence of the matrix elements of the twist-2 operators matches 
the scale dependence of the perturbatively calculated Wilson
coefficients at most at large scales, so that we may expect large 
non-perturbative effects.
A lattice calculation of the Wilson coefficients would also save us
to renormalize the lattice operators. 

We consider DIS off a quark target. This is sufficient for the
calculation of the `leading' Wilson coefficients. The OPE reads
\begin{eqnarray}
W_{\mu\nu}(p,q) \!\!&\equiv&\!\! \langle p| J_\mu(q) J_\nu(-q)
  |p\rangle \nonumber + \mbox{seagull}\\
\!\!&=&\!\! \sum_n C_{n,\mu\nu}(\Lambda^{-1}\,q)\, O_n(\Lambda^{-1}\,p),
\end{eqnarray}
where $|p\rangle$ is an off-shell quark state of momentum $p$
which is struck by a photon of momentum $q$, and $\Lambda$ is a
scale parameter. The $C_{n,\mu\nu}(\Lambda^{-1}\,q)$
are the Wilson coefficients, and $O_n(\Lambda^{-1}\,p)$ are the forward
matrix elements of the operators
\begin{equation}
{\cal O}_{\mu_1 \dots \mu_n} = \bar{\psi}\Gamma_{\mu_1}D_{\mu_2} 
\dots D_{\mu_n} \psi,
\label{ops}
\end{equation}
where $\Gamma$ is a matrix in Dirac space. In the following 
we will distinguish between bare and renormalized quantities. In the 
former $\Lambda = a^{-1}$, where $a$ is the lattice spacing. In the 
latter $\Lambda = \mu$, where $\mu$ is the renormalization scale.  

The Wilson coefficients are independent of the target. 
Let us denote the nucleon quantities by the superscript 
${\scriptstyle N}$. 
So if we multiply the nucleon matrix elements of the operators (\ref{ops})
by the corresponding Wilson coefficients, 
we get the nucleon structure functions: 
\begin{equation}
W_{\mu\nu}^{(N)}(p,q) = \sum_n C_{n,\mu\nu}(\Lambda^{-1}\,q)\, 
O_n^{(N)}(\Lambda^{-1}\,p).
\label{nucleon}
\end{equation}
The r.h.s. of (\ref{nucleon}), being independent of $\Lambda$, can be
written in different ways:
\begin{equation}
\begin{array}{l}
C_{n,\mu\nu}(a\,q)\, O_n^{(N)}(a\,p) \\ 
= C_{n,\mu\nu}(\mu^{-1}\,q)\, O_n^{(N)}(\mu^{-1}\,p) \\ 
= C_{n,\mu\nu}(\mu^{-1}\,q)\, Z_n(\mu\,a)\, O_n^{(N)}(a\,p),
\end{array}
\end{equation}
where $Z_n(\mu\,a)$ is the renormalization constant of the
lattice operator~\cite{renorm,renorm2}. As a result, the
renormalized and bare Wilson coefficients are related by
\begin{equation}
C_{n,\mu\nu}(\mu^{-1}\,q) = Z_n^{-1}(\mu\,a)\, C_{n,\mu\nu}(a\,q).
\end{equation}
The nucleon structure functions are then most conveniently computed
from the product of bare Wilson coefficients and lattice nucleon
matrix elements. A further advantage of this approach is that operator
mixing with higher-twist operators is automatically included.
 
\begin{figure}[htb]
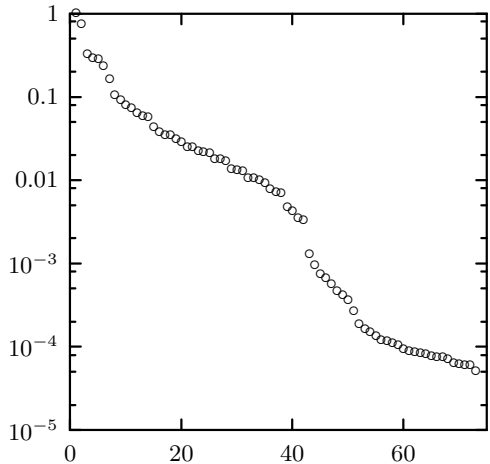

\small
\begin{center}
\input SV.eepic
\vspace*{-0.9cm}
\caption{The eigenvalues $w_n$  
  against $n$, normalized so that the largest value is
  one.}
\vspace*{-0.9cm}
\label{SVplot}
\end{center}
\end{figure}

The plan is now to compute the tensor $W_{\mu\nu}$ and the operator
matrix elements $O_n$ for a quark in the Landau gauge, and to extract the
Wilson coefficients from this information. 

\section{THE METHOD}

We consider a system of quark momenta $p_m, \, 1 \leq m \leq M$ with the
photon momentum being fixed. The Wilson coefficients are independent of
$p_m$, so the problem is to solve the $N \times M$ system of equations:
\begin{equation}
\left(
\begin{array}{ccc}
 O_{1}^{p_1}  &  \hspace*{-0.27cm}\cdots\hspace{-0.27cm} & 
 O_{N}^{p_1}   \\
\vdots       & \hspace*{-0.27cm} \hspace{-0.27cm} & 
\vdots   \\
 O_{1}^{p_M}  & \hspace*{-0.27cm}\cdots\hspace*{-0.27cm} & 
 O_{N}^{p_M}   \\
\end{array}
\right)
\left(
\begin{array}{ccc}
C_{1}  \\
\vdots \\
C_{N}  \\
\end{array}
\right)
=
\left(
\begin{array}{ccc}
W^{p_1}  \\
\vdots    \\
W^{p_M}  \\
\end{array}
\right),
\label{matrix}
\end{equation}
\begin{figure}[htb]
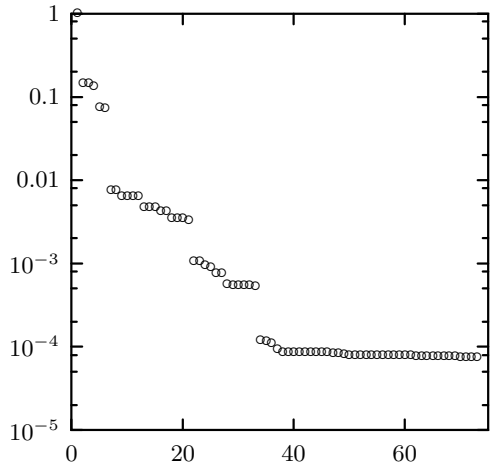

\small
\begin{center}
\input chi.eepic
\vspace*{-0.9cm}
\caption{The residual error $R^2$ against $n$.}
\vspace*{-0.9cm}
\label{chi}
\end{center}
\end{figure}
where $O_n^{p_m}$ are the ensemble averaged (amputated) matrix
elements of the 
operators (\ref{ops}) between quark states of momentum $p_m$,
$W^{p_m}$ are the corresponding elements of the hadronic tensor, and
the $C_n$ are the Wilson coefficients we are looking for. The label
$n$ runs over the various operators. 
Note that Lorentz and Dirac indices have been omitted.
We shall take all operators with
up to three covariant derivatives into account. Since we do not symmetrize
over the indices $\mu_1, \mu_2, \cdots$, this includes higher-twist
operators as well. 

We write (\ref{matrix}) as
\begin{equation}
O C = W.
\end{equation}
To compute $C$, we apply a singular value decomposition,
which is the standard method of solving overdetermined 
equations~\cite{numrec}.
We write 
\begin{equation}
O = U w V^T,
\label{o}
\end{equation}
where $U$ is a column-orthonormal $M\times N$ matrix, $w$ is a diagonal
$N\times N$ matrix, $w = \mbox{diag}(w_n)$, with positive real eigenvalues 
$w_n,\, 1 \leq n \leq N$ arranged in descending order, and
$V^T$ is the transpose of a column-orthonormal $N\times
N$ matrix. The solution $C$ is obtained by applying
\begin{equation}
V \,\mbox{diag}(1/w_n)\, U^T
\end{equation}
to $W$. To find the vector $C$ of smallest length, we replace $1/w_n$ by zero
if $w_n$ is dominated by noise. 
\begin{figure}[htb]
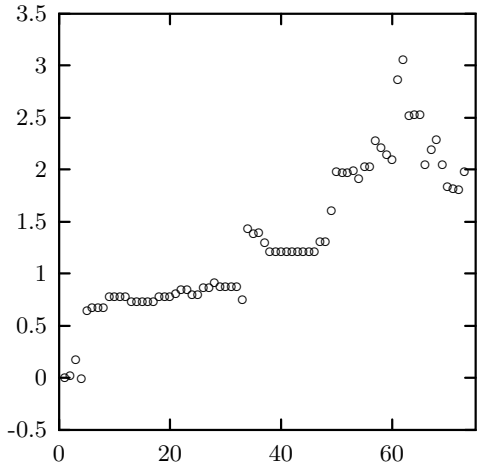

\small
\begin{center}
\input OP_35.eepic
\vspace*{-0.9cm}
\caption{The Wilson coefficient $c_x(a\,q)$ of the operator $X =
  \bar{\psi}\gamma_{\{i} D_{j\}}\psi$ with $i\neq j$ against $n$.}
\vspace*{-0.9cm}
\label{OP_35}
\end{center}
\end{figure}

\section{RESULTS}

The calculations are done with standard Wilson fermions on $24^3 48$
lattices at $\beta = 6.2$ and $\kappa = 0.1489$. So far we have looked
at two configurations. We employed 71 different quark momenta. For the
photon momentum we took 
\begin{equation}
q^2 = (\pi/2a)^2 = 17.4 \; \mbox{GeV}^2.
\label{photon}
\end{equation}

In Fig.~1 we show the eigenvalues $w_n$ of $w$.
We see a sharp decrease between $n \approx 40$ and 50 by approximately
a factor of 30, and the $w_n$ for larger $n$ are probably dominated by
noise. 
In Fig.~2 we have plotted the
residual error $R^2 = |W-OC|^2/|W|^2$ that results from setting
$1/w_m = 0,\, \forall\, m \geq n$, as a function of $n$. The error
drops sharply until $n$ reaches $\approx 40$ and then stays constant,
saying that if one makes a fit with too few parameters one does
not get a good fit, and if one fits with too many one `fits to 
the noise'.
So we must truncate the system at $n \approx 40 - 50$.

The commonly used Wilson coefficients (bare of all kinematical factors) are
obtained by
\begin{equation}
c_n(a\,q) = C_{n, \mu\nu}(a\,q)/C_{n, \mu\nu}^{\scriptstyle
 {\rm tree}}(a\,q), 
\end{equation}
where $C_{n, \mu\nu}^{\scriptstyle {\rm tree}}(a\,q)$ is the tree 
value. In Fig.~3 we show the Wilson coefficient
$c_x(a\,q)$~\cite{structure} of the
operator $X=\bar{\psi}\gamma_{\{i} D_{j\}}\psi$ with $i\neq j$, again 
for various
degrees of truncation. We find a plateau in the region where the
eigenvalues $w_n$ drop to zero, so that it does not matter where exactly we
truncate the system. We have chosen $n=42$. Beyond the plateau the
errors on $w_n$ get large. 

\begin{figure}[htb]
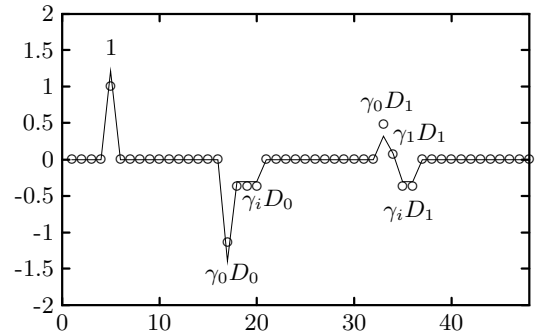

\small
\begin{center}
\vspace*{0.16cm}
\input OP_1-48.eepic
\vspace*{-0.4cm}
\caption{The Wilson coefficients $c_n(a\,q)$ of the first 48
  operators. The open circles are our Monte-Carlo results. The solid
  line is the tree-level result. The horizontal axis labels the
  operators, including the 16 different $\Gamma$ matrices.}
\label{OP_1-48}
\vspace*{-0.9cm}
\end{center}
\end{figure}

As a first step we have verified the OPE. In Fig.~4 we show the Wilson
coefficients $c_n(a\,q)$ of the lower dimensional operators. We
compare our results with the tree-level predictions.
We find that the expected structure is very well reproduced.
In particular, the Wilson coefficients which should be zero by
symmetry or other arguments are actually found to be zero. 

As a first application of our method we have determined the Wilson
coefficient of the operator $X$ already shown in
Fig.~3. This operator gives the first (non-trivial) moment of the
unpolarized structure function~\cite{structure}, $\langle x \rangle$.
We find (at $q^2 = 17.4 \, \mbox{GeV}^2$, cf. (\ref{photon}))
$c_x(a\,q) \approx 1.2$.
If we divide this number by the corresponding renormalization constant,
$Z_x^{\overline{MS}}(\mu\,a)$, as given in~\cite{renorm}, we
obtain 
\begin{equation}
c_x^{\overline{MS}}(\mu^{-1}\,q)|_{\mu^2=q^2} \approx 1.2.
\end{equation} 
This is to be compared with the 2-loop result of
$c_x^{\overline{MS}}(\mu^2=q^2) = 1.01$.


\end{document}

%% file: SV.eepic
\setlength{\unitlength}{0.2000pt}
\begin{picture}(1500,900)(0,0)
\thicklines \path(218,67)(238,67)
\thicklines \path(1006,67)(986,67)
\put(196,67){\makebox(0,0)[r]{$10^{-5}$}}
\thicklines \path(218,114)(228,114)
\thicklines \path(1006,114)(996,114)
\thicklines \path(218,177)(228,177)
\thicklines \path(1006,177)(996,177)
\thicklines \path(218,209)(228,209)
\thicklines \path(1006,209)(996,209)
\thicklines \path(218,225)(238,225)
\thicklines \path(1006,225)(986,225)
\put(196,225){\makebox(0,0)[r]{$10^{-4}$}}
\thicklines \path(218,272)(228,272)
\thicklines \path(1006,272)(996,272)
\thicklines \path(218,335)(228,335)
\thicklines \path(1006,335)(996,335)
\thicklines \path(218,367)(228,367)
\thicklines \path(1006,367)(996,367)
\thicklines \path(218,382)(238,382)
\thicklines \path(1006,382)(986,382)
\put(196,382){\makebox(0,0)[r]{$10^{-3}$}}
\thicklines \path(218,430)(228,430)
\thicklines \path(1006,430)(996,430)
\thicklines \path(218,492)(228,492)
\thicklines \path(1006,492)(996,492)
\thicklines \path(218,525)(228,525)
\thicklines \path(1006,525)(996,525)
\thicklines \path(218,540)(238,540)
\thicklines \path(1006,540)(986,540)
\put(196,540){\makebox(0,0)[r]{0.01}}
\thicklines \path(218,587)(228,587)
\thicklines \path(1006,587)(996,587)
\thicklines \path(218,650)(228,650)
\thicklines \path(1006,650)(996,650)
\thicklines \path(218,682)(228,682)
\thicklines \path(1006,682)(996,682)
\thicklines \path(218,697)(238,697)
\thicklines \path(1006,697)(986,697)
\put(196,697){\makebox(0,0)[r]{0.1}}
\thicklines \path(218,745)(228,745)
\thicklines \path(1006,745)(996,745)
\thicklines \path(218,808)(228,808)
\thicklines \path(1006,808)(996,808)
\thicklines \path(218,840)(228,840)
\thicklines \path(1006,840)(996,840)
\thicklines \path(218,855)(238,855)
\thicklines \path(1006,855)(986,855)
\put(196,855){\makebox(0,0)[r]{1}}
\thicklines \path(218,67)(218,87)
\thicklines \path(218,855)(218,835)
\put(218,22){\makebox(0,0){0}}
\thicklines \path(428,67)(428,87)
\thicklines \path(428,855)(428,835)
\put(428,22){\makebox(0,0){20}}
\thicklines \path(638,67)(638,87)
\thicklines \path(638,855)(638,835)
\put(638,22){\makebox(0,0){40}}
\thicklines \path(848,67)(848,87)
\thicklines \path(848,855)(848,835)
\put(848,22){\makebox(0,0){60}}
\thicklines \path(218,67)(1006,67)(1006,855)(218,855)(218,67)
\put(229,855){\raisebox{-0.0pt}{\makebox(0,0){\footnotesize $\circ$}}}
\put(239,834){\raisebox{-0.0pt}{\makebox(0,0){\footnotesize $\circ$}}}
\put(250,778){\raisebox{-0.0pt}{\makebox(0,0){\footnotesize $\circ$}}}
\put(260,770){\raisebox{-0.0pt}{\makebox(0,0){\footnotesize $\circ$}}}
\put(271,769){\raisebox{-0.0pt}{\makebox(0,0){\footnotesize $\circ$}}}
\put(281,755){\raisebox{-0.0pt}{\makebox(0,0){\footnotesize $\circ$}}}
\put(292,730){\raisebox{-0.0pt}{\makebox(0,0){\footnotesize $\circ$}}}
\put(302,700){\raisebox{-0.0pt}{\makebox(0,0){\footnotesize $\circ$}}}
\put(313,690){\raisebox{-0.0pt}{\makebox(0,0){\footnotesize $\circ$}}}
\put(323,681){\raisebox{-0.0pt}{\makebox(0,0){\footnotesize $\circ$}}}
\put(334,675){\raisebox{-0.0pt}{\makebox(0,0){\footnotesize $\circ$}}}
\put(344,666){\raisebox{-0.0pt}{\makebox(0,0){\footnotesize $\circ$}}}
\put(355,661){\raisebox{-0.0pt}{\makebox(0,0){\footnotesize $\circ$}}}
\put(365,659){\raisebox{-0.0pt}{\makebox(0,0){\footnotesize $\circ$}}}
\put(376,640){\raisebox{-0.0pt}{\makebox(0,0){\footnotesize $\circ$}}}
\put(386,630){\raisebox{-0.0pt}{\makebox(0,0){\footnotesize $\circ$}}}
\put(397,625){\raisebox{-0.0pt}{\makebox(0,0){\footnotesize $\circ$}}}
\put(407,625){\raisebox{-0.0pt}{\makebox(0,0){\footnotesize $\circ$}}}
\put(418,618){\raisebox{-0.0pt}{\makebox(0,0){\footnotesize $\circ$}}}
\put(428,612){\raisebox{-0.0pt}{\makebox(0,0){\footnotesize $\circ$}}}
\put(439,603){\raisebox{-0.0pt}{\makebox(0,0){\footnotesize $\circ$}}}
\put(449,602){\raisebox{-0.0pt}{\makebox(0,0){\footnotesize $\circ$}}}
\put(460,594){\raisebox{-0.0pt}{\makebox(0,0){\footnotesize $\circ$}}}
\put(470,593){\raisebox{-0.0pt}{\makebox(0,0){\footnotesize $\circ$}}}
\put(481,590){\raisebox{-0.0pt}{\makebox(0,0){\footnotesize $\circ$}}}
\put(491,579){\raisebox{-0.0pt}{\makebox(0,0){\footnotesize $\circ$}}}
\put(502,579){\raisebox{-0.0pt}{\makebox(0,0){\footnotesize $\circ$}}}
\put(512,575){\raisebox{-0.0pt}{\makebox(0,0){\footnotesize $\circ$}}}
\put(523,561){\raisebox{-0.0pt}{\makebox(0,0){\footnotesize $\circ$}}}
\put(533,559){\raisebox{-0.0pt}{\makebox(0,0){\footnotesize $\circ$}}}
\put(544,557){\raisebox{-0.0pt}{\makebox(0,0){\footnotesize $\circ$}}}
\put(554,544){\raisebox{-0.0pt}{\makebox(0,0){\footnotesize $\circ$}}}
\put(565,543){\raisebox{-0.0pt}{\makebox(0,0){\footnotesize $\circ$}}}
\put(575,540){\raisebox{-0.0pt}{\makebox(0,0){\footnotesize $\circ$}}}
\put(586,534){\raisebox{-0.0pt}{\makebox(0,0){\footnotesize $\circ$}}}
\put(596,522){\raisebox{-0.0pt}{\makebox(0,0){\footnotesize $\circ$}}}
\put(607,517){\raisebox{-0.0pt}{\makebox(0,0){\footnotesize $\circ$}}}
\put(617,514){\raisebox{-0.0pt}{\makebox(0,0){\footnotesize $\circ$}}}
\put(628,489){\raisebox{-0.0pt}{\makebox(0,0){\footnotesize $\circ$}}}
\put(638,481){\raisebox{-0.0pt}{\makebox(0,0){\footnotesize $\circ$}}}
\put(649,468){\raisebox{-0.0pt}{\makebox(0,0){\footnotesize $\circ$}}}
\put(659,463){\raisebox{-0.0pt}{\makebox(0,0){\footnotesize $\circ$}}}
\put(670,400){\raisebox{-0.0pt}{\makebox(0,0){\footnotesize $\circ$}}}
\put(680,378){\raisebox{-0.0pt}{\makebox(0,0){\footnotesize $\circ$}}}
\put(691,361){\raisebox{-0.0pt}{\makebox(0,0){\footnotesize $\circ$}}}
\put(701,354){\raisebox{-0.0pt}{\makebox(0,0){\footnotesize $\circ$}}}
\put(712,342){\raisebox{-0.0pt}{\makebox(0,0){\footnotesize $\circ$}}}
\put(722,330){\raisebox{-0.0pt}{\makebox(0,0){\footnotesize $\circ$}}}
\put(733,322){\raisebox{-0.0pt}{\makebox(0,0){\footnotesize $\circ$}}}
\put(743,313){\raisebox{-0.0pt}{\makebox(0,0){\footnotesize $\circ$}}}
\put(754,292){\raisebox{-0.0pt}{\makebox(0,0){\footnotesize $\circ$}}}
\put(764,267){\raisebox{-0.0pt}{\makebox(0,0){\footnotesize $\circ$}}}
\put(775,258){\raisebox{-0.0pt}{\makebox(0,0){\footnotesize $\circ$}}}
\put(785,252){\raisebox{-0.0pt}{\makebox(0,0){\footnotesize $\circ$}}}
\put(796,244){\raisebox{-0.0pt}{\makebox(0,0){\footnotesize $\circ$}}}
\put(806,236){\raisebox{-0.0pt}{\makebox(0,0){\footnotesize $\circ$}}}
\put(817,235){\raisebox{-0.0pt}{\makebox(0,0){\footnotesize $\circ$}}}
\put(827,230){\raisebox{-0.0pt}{\makebox(0,0){\footnotesize $\circ$}}}
\put(838,227){\raisebox{-0.0pt}{\makebox(0,0){\footnotesize $\circ$}}}
\put(848,220){\raisebox{-0.0pt}{\makebox(0,0){\footnotesize $\circ$}}}
\put(859,216){\raisebox{-0.0pt}{\makebox(0,0){\footnotesize $\circ$}}}
\put(869,214){\raisebox{-0.0pt}{\makebox(0,0){\footnotesize $\circ$}}}
\put(880,212){\raisebox{-0.0pt}{\makebox(0,0){\footnotesize $\circ$}}}
\put(890,210){\raisebox{-0.0pt}{\makebox(0,0){\footnotesize $\circ$}}}
\put(901,207){\raisebox{-0.0pt}{\makebox(0,0){\footnotesize $\circ$}}}
\put(911,205){\raisebox{-0.0pt}{\makebox(0,0){\footnotesize $\circ$}}}
\put(922,204){\raisebox{-0.0pt}{\makebox(0,0){\footnotesize $\circ$}}}
\put(932,201){\raisebox{-0.0pt}{\makebox(0,0){\footnotesize $\circ$}}}
\put(943,193){\raisebox{-0.0pt}{\makebox(0,0){\footnotesize $\circ$}}}
\put(953,191){\raisebox{-0.0pt}{\makebox(0,0){\footnotesize $\circ$}}}
\put(964,190){\raisebox{-0.0pt}{\makebox(0,0){\footnotesize $\circ$}}}
\put(974,190){\raisebox{-0.0pt}{\makebox(0,0){\footnotesize $\circ$}}}
\put(985,178){\raisebox{-0.0pt}{\makebox(0,0){\footnotesize $\circ$}}}
\end{picture}

%% file: chi.eepic
\setlength{\unitlength}{0.2000pt}
\begin{picture}(1500,900)(0,0)
\thicklines \path(218,67)(238,67)
\thicklines \path(1006,67)(986,67)
\put(196,67){\makebox(0,0)[r]{$10^{-5}$}}
\thicklines \path(218,114)(228,114)
\thicklines \path(1006,114)(996,114)
\thicklines \path(218,177)(228,177)
\thicklines \path(1006,177)(996,177)
\thicklines \path(218,209)(228,209)
\thicklines \path(1006,209)(996,209)
\thicklines \path(218,225)(238,225)
\thicklines \path(1006,225)(986,225)
\put(196,225){\makebox(0,0)[r]{$10^{-4}$}}
\thicklines \path(218,272)(228,272)
\thicklines \path(1006,272)(996,272)
\thicklines \path(218,335)(228,335)
\thicklines \path(1006,335)(996,335)
\thicklines \path(218,367)(228,367)
\thicklines \path(1006,367)(996,367)
\thicklines \path(218,382)(238,382)
\thicklines \path(1006,382)(986,382)
\put(196,382){\makebox(0,0)[r]{$10^{-3}$}}
\thicklines \path(218,430)(228,430)
\thicklines \path(1006,430)(996,430)
\thicklines \path(218,492)(228,492)
\thicklines \path(1006,492)(996,492)
\thicklines \path(218,525)(228,525)
\thicklines \path(1006,525)(996,525)
\thicklines \path(218,540)(238,540)
\thicklines \path(1006,540)(986,540)
\put(196,540){\makebox(0,0)[r]{0.01}}
\thicklines \path(218,587)(228,587)
\thicklines \path(1006,587)(996,587)
\thicklines \path(218,650)(228,650)
\thicklines \path(1006,650)(996,650)
\thicklines \path(218,682)(228,682)
\thicklines \path(1006,682)(996,682)
\thicklines \path(218,697)(238,697)
\thicklines \path(1006,697)(986,697)
\put(196,697){\makebox(0,0)[r]{0.1}}
\thicklines \path(218,745)(228,745)
\thicklines \path(1006,745)(996,745)
\thicklines \path(218,808)(228,808)
\thicklines \path(1006,808)(996,808)
\thicklines \path(218,840)(228,840)
\thicklines \path(1006,840)(996,840)
\thicklines \path(218,855)(238,855)
\thicklines \path(1006,855)(986,855)
\put(196,855){\makebox(0,0)[r]{1}}
\thicklines \path(218,67)(218,87)
\thicklines \path(218,855)(218,835)
\put(218,22){\makebox(0,0){0}}
\thicklines \path(428,67)(428,87)
\thicklines \path(428,855)(428,835)
\put(428,22){\makebox(0,0){20}}
\thicklines \path(638,67)(638,87)
\thicklines \path(638,855)(638,835)
\put(638,22){\makebox(0,0){40}}
\thicklines \path(848,67)(848,87)
\thicklines \path(848,855)(848,835)
\put(848,22){\makebox(0,0){60}}
\thicklines \path(218,67)(1006,67)(1006,855)(218,855)(218,67)
\put(229,855){\raisebox{0.0pt}{\makebox(0,0){\footnotesize $\circ$}}}
\put(239,723){\raisebox{0.0pt}{\makebox(0,0){\footnotesize $\circ$}}}
\put(250,723){\raisebox{0.0pt}{\makebox(0,0){\footnotesize $\circ$}}}
\put(260,717){\raisebox{0.0pt}{\makebox(0,0){\footnotesize $\circ$}}}
\put(271,677){\raisebox{0.0pt}{\makebox(0,0){\footnotesize $\circ$}}}
\put(281,676){\raisebox{0.0pt}{\makebox(0,0){\footnotesize $\circ$}}}
\put(292,520){\raisebox{0.0pt}{\makebox(0,0){\footnotesize $\circ$}}}
\put(302,520){\raisebox{0.0pt}{\makebox(0,0){\footnotesize $\circ$}}}
\put(313,510){\raisebox{0.0pt}{\makebox(0,0){\footnotesize $\circ$}}}
\put(323,510){\raisebox{0.0pt}{\makebox(0,0){\footnotesize $\circ$}}}
\put(334,510){\raisebox{0.0pt}{\makebox(0,0){\footnotesize $\circ$}}}
\put(344,510){\raisebox{0.0pt}{\makebox(0,0){\footnotesize $\circ$}}}
\put(355,489){\raisebox{0.0pt}{\makebox(0,0){\footnotesize $\circ$}}}
\put(365,489){\raisebox{0.0pt}{\makebox(0,0){\footnotesize $\circ$}}}
\put(376,489){\raisebox{0.0pt}{\makebox(0,0){\footnotesize $\circ$}}}
\put(386,481){\raisebox{0.0pt}{\makebox(0,0){\footnotesize $\circ$}}}
\put(397,481){\raisebox{0.0pt}{\makebox(0,0){\footnotesize $\circ$}}}
\put(407,468){\raisebox{0.0pt}{\makebox(0,0){\footnotesize $\circ$}}}
\put(418,468){\raisebox{0.0pt}{\makebox(0,0){\footnotesize $\circ$}}}
\put(428,468){\raisebox{0.0pt}{\makebox(0,0){\footnotesize $\circ$}}}
\put(439,463){\raisebox{0.0pt}{\makebox(0,0){\footnotesize $\circ$}}}
\put(449,386){\raisebox{0.0pt}{\makebox(0,0){\footnotesize $\circ$}}}
\put(460,386){\raisebox{0.0pt}{\makebox(0,0){\footnotesize $\circ$}}}
\put(470,378){\raisebox{0.0pt}{\makebox(0,0){\footnotesize $\circ$}}}
\put(481,374){\raisebox{0.0pt}{\makebox(0,0){\footnotesize $\circ$}}}
\put(491,364){\raisebox{0.0pt}{\makebox(0,0){\footnotesize $\circ$}}}
\put(502,364){\raisebox{0.0pt}{\makebox(0,0){\footnotesize $\circ$}}}
\put(512,342){\raisebox{0.0pt}{\makebox(0,0){\footnotesize $\circ$}}}
\put(523,340){\raisebox{0.0pt}{\makebox(0,0){\footnotesize $\circ$}}}
\put(533,340){\raisebox{0.0pt}{\makebox(0,0){\footnotesize $\circ$}}}
\put(544,340){\raisebox{0.0pt}{\makebox(0,0){\footnotesize $\circ$}}}
\put(554,340){\raisebox{0.0pt}{\makebox(0,0){\footnotesize $\circ$}}}
\put(565,339){\raisebox{0.0pt}{\makebox(0,0){\footnotesize $\circ$}}}
\put(575,236){\raisebox{0.0pt}{\makebox(0,0){\footnotesize $\circ$}}}
\put(586,235){\raisebox{0.0pt}{\makebox(0,0){\footnotesize $\circ$}}}
\put(596,232){\raisebox{0.0pt}{\makebox(0,0){\footnotesize $\circ$}}}
\put(607,220){\raisebox{0.0pt}{\makebox(0,0){\footnotesize $\circ$}}}
\put(617,214){\raisebox{0.0pt}{\makebox(0,0){\footnotesize $\circ$}}}
\put(628,214){\raisebox{0.0pt}{\makebox(0,0){\footnotesize $\circ$}}}
\put(638,214){\raisebox{0.0pt}{\makebox(0,0){\footnotesize $\circ$}}}
\put(649,214){\raisebox{0.0pt}{\makebox(0,0){\footnotesize $\circ$}}}
\put(659,214){\raisebox{0.0pt}{\makebox(0,0){\footnotesize $\circ$}}}
\put(670,214){\raisebox{0.0pt}{\makebox(0,0){\footnotesize $\circ$}}}
\put(680,214){\raisebox{0.0pt}{\makebox(0,0){\footnotesize $\circ$}}}
\put(691,214){\raisebox{0.0pt}{\makebox(0,0){\footnotesize $\circ$}}}
\put(701,214){\raisebox{0.0pt}{\makebox(0,0){\footnotesize $\circ$}}}
\put(712,212){\raisebox{0.0pt}{\makebox(0,0){\footnotesize $\circ$}}}
\put(722,212){\raisebox{0.0pt}{\makebox(0,0){\footnotesize $\circ$}}}
\put(733,210){\raisebox{0.0pt}{\makebox(0,0){\footnotesize $\circ$}}}
\put(743,208){\raisebox{0.0pt}{\makebox(0,0){\footnotesize $\circ$}}}
\put(754,208){\raisebox{0.0pt}{\makebox(0,0){\footnotesize $\circ$}}}
\put(764,208){\raisebox{0.0pt}{\makebox(0,0){\footnotesize $\circ$}}}
\put(775,208){\raisebox{0.0pt}{\makebox(0,0){\footnotesize $\circ$}}}
\put(785,208){\raisebox{0.0pt}{\makebox(0,0){\footnotesize $\circ$}}}
\put(796,208){\raisebox{0.0pt}{\makebox(0,0){\footnotesize $\circ$}}}
\put(806,208){\raisebox{0.0pt}{\makebox(0,0){\footnotesize $\circ$}}}
\put(817,208){\raisebox{0.0pt}{\makebox(0,0){\footnotesize $\circ$}}}
\put(827,208){\raisebox{0.0pt}{\makebox(0,0){\footnotesize $\circ$}}}
\put(838,208){\raisebox{0.0pt}{\makebox(0,0){\footnotesize $\circ$}}}
\put(848,208){\raisebox{0.0pt}{\makebox(0,0){\footnotesize $\circ$}}}
\put(859,208){\raisebox{0.0pt}{\makebox(0,0){\footnotesize $\circ$}}}
\put(869,207){\raisebox{0.0pt}{\makebox(0,0){\footnotesize $\circ$}}}
\put(880,207){\raisebox{0.0pt}{\makebox(0,0){\footnotesize $\circ$}}}
\put(890,207){\raisebox{0.0pt}{\makebox(0,0){\footnotesize $\circ$}}}
\put(901,207){\raisebox{0.0pt}{\makebox(0,0){\footnotesize $\circ$}}}
\put(911,207){\raisebox{0.0pt}{\makebox(0,0){\footnotesize $\circ$}}}
\put(922,207){\raisebox{0.0pt}{\makebox(0,0){\footnotesize $\circ$}}}
\put(932,207){\raisebox{0.0pt}{\makebox(0,0){\footnotesize $\circ$}}}
\put(943,206){\raisebox{0.0pt}{\makebox(0,0){\footnotesize $\circ$}}}
\put(953,205){\raisebox{0.0pt}{\makebox(0,0){\footnotesize $\circ$}}}
\put(964,205){\raisebox{0.0pt}{\makebox(0,0){\footnotesize $\circ$}}}
\put(974,205){\raisebox{0.0pt}{\makebox(0,0){\footnotesize $\circ$}}}
\put(985,204){\raisebox{0.0pt}{\makebox(0,0){\footnotesize $\circ$}}}
\end{picture}

%% file: OP_35.eepic
\setlength{\unitlength}{0.2000pt}
\begin{picture}(1500,900)(0,0)
\thicklines \path(174,67)(194,67)
\thicklines \path(962,67)(942,67)
\put(152,67){\makebox(0,0)[r]{-0.5}}
\thicklines \path(174,166)(194,166)
\thicklines \path(962,166)(942,166)
\put(152,166){\makebox(0,0)[r]{0}}
\thicklines \path(174,264)(194,264)
\thicklines \path(962,264)(942,264)
\put(152,264){\makebox(0,0)[r]{0.5}}
\thicklines \path(174,363)(194,363)
\thicklines \path(962,363)(942,363)
\put(152,363){\makebox(0,0)[r]{1}}
\thicklines \path(174,461)(194,461)
\thicklines \path(962,461)(942,461)
\put(152,461){\makebox(0,0)[r]{1.5}}
\thicklines \path(174,560)(194,560)
\thicklines \path(962,560)(942,560)
\put(152,560){\makebox(0,0)[r]{2}}
\thicklines \path(174,658)(194,658)
\thicklines \path(962,658)(942,658)
\put(152,658){\makebox(0,0)[r]{2.5}}
\thicklines \path(174,757)(194,757)
\thicklines \path(962,757)(942,757)
\put(152,757){\makebox(0,0)[r]{3}}
\thicklines \path(174,855)(194,855)
\thicklines \path(962,855)(942,855)
\put(152,855){\makebox(0,0)[r]{3.5}}
\thicklines \path(174,67)(174,87)
\thicklines \path(174,855)(174,835)
\put(174,22){\makebox(0,0){0}}
\thicklines \path(384,67)(384,87)
\thicklines \path(384,855)(384,835)
\put(384,22){\makebox(0,0){20}}
\thicklines \path(594,67)(594,87)
\thicklines \path(594,855)(594,835)
\put(594,22){\makebox(0,0){40}}
\thicklines \path(804,67)(804,87)
\thicklines \path(804,855)(804,835)
\put(804,22){\makebox(0,0){60}}
\thicklines \path(174,67)(962,67)(962,855)(174,855)(174,67)
\put(185,165){\raisebox{0.0pt}{\makebox(0,0){\footnotesize $\circ$}}}
\put(195,169){\raisebox{0.0pt}{\makebox(0,0){\footnotesize $\circ$}}}
\put(206,199){\raisebox{0.0pt}{\makebox(0,0){\footnotesize $\circ$}}}
\put(216,163){\raisebox{0.0pt}{\makebox(0,0){\footnotesize $\circ$}}}
\put(227,291){\raisebox{0.0pt}{\makebox(0,0){\footnotesize $\circ$}}}
\put(237,297){\raisebox{0.0pt}{\makebox(0,0){\footnotesize $\circ$}}}
\put(248,298){\raisebox{0.0pt}{\makebox(0,0){\footnotesize $\circ$}}}
\put(258,298){\raisebox{0.0pt}{\makebox(0,0){\footnotesize $\circ$}}}
\put(269,317){\raisebox{0.0pt}{\makebox(0,0){\footnotesize $\circ$}}}
\put(279,317){\raisebox{0.0pt}{\makebox(0,0){\footnotesize $\circ$}}}
\put(290,317){\raisebox{0.0pt}{\makebox(0,0){\footnotesize $\circ$}}}
\put(300,317){\raisebox{0.0pt}{\makebox(0,0){\footnotesize $\circ$}}}
\put(311,308){\raisebox{0.0pt}{\makebox(0,0){\footnotesize $\circ$}}}
\put(321,308){\raisebox{0.0pt}{\makebox(0,0){\footnotesize $\circ$}}}
\put(332,308){\raisebox{0.0pt}{\makebox(0,0){\footnotesize $\circ$}}}
\put(342,308){\raisebox{0.0pt}{\makebox(0,0){\footnotesize $\circ$}}}
\put(353,308){\raisebox{0.0pt}{\makebox(0,0){\footnotesize $\circ$}}}
\put(363,318){\raisebox{0.0pt}{\makebox(0,0){\footnotesize $\circ$}}}
\put(374,318){\raisebox{0.0pt}{\makebox(0,0){\footnotesize $\circ$}}}
\put(384,318){\raisebox{0.0pt}{\makebox(0,0){\footnotesize $\circ$}}}
\put(395,324){\raisebox{0.0pt}{\makebox(0,0){\footnotesize $\circ$}}}
\put(405,332){\raisebox{0.0pt}{\makebox(0,0){\footnotesize $\circ$}}}
\put(416,332){\raisebox{0.0pt}{\makebox(0,0){\footnotesize $\circ$}}}
\put(426,322){\raisebox{0.0pt}{\makebox(0,0){\footnotesize $\circ$}}}
\put(437,321){\raisebox{0.0pt}{\makebox(0,0){\footnotesize $\circ$}}}
\put(447,335){\raisebox{0.0pt}{\makebox(0,0){\footnotesize $\circ$}}}
\put(458,335){\raisebox{0.0pt}{\makebox(0,0){\footnotesize $\circ$}}}
\put(468,345){\raisebox{0.0pt}{\makebox(0,0){\footnotesize $\circ$}}}
\put(479,338){\raisebox{0.0pt}{\makebox(0,0){\footnotesize $\circ$}}}
\put(489,338){\raisebox{0.0pt}{\makebox(0,0){\footnotesize $\circ$}}}
\put(500,336){\raisebox{0.0pt}{\makebox(0,0){\footnotesize $\circ$}}}
\put(510,336){\raisebox{0.0pt}{\makebox(0,0){\footnotesize $\circ$}}}
\put(521,312){\raisebox{0.0pt}{\makebox(0,0){\footnotesize $\circ$}}}
\put(531,447){\raisebox{0.0pt}{\makebox(0,0){\footnotesize $\circ$}}}
\put(542,438){\raisebox{0.0pt}{\makebox(0,0){\footnotesize $\circ$}}}
\put(552,439){\raisebox{0.0pt}{\makebox(0,0){\footnotesize $\circ$}}}
\put(563,420){\raisebox{0.0pt}{\makebox(0,0){\footnotesize $\circ$}}}
\put(573,404){\raisebox{0.0pt}{\makebox(0,0){\footnotesize $\circ$}}}
\put(584,404){\raisebox{0.0pt}{\makebox(0,0){\footnotesize $\circ$}}}
\put(594,404){\raisebox{0.0pt}{\makebox(0,0){\footnotesize $\circ$}}}
\put(605,404){\raisebox{0.0pt}{\makebox(0,0){\footnotesize $\circ$}}}
\put(615,404){\raisebox{0.0pt}{\makebox(0,0){\footnotesize $\circ$}}}
\put(626,404){\raisebox{0.0pt}{\makebox(0,0){\footnotesize $\circ$}}}
\put(636,404){\raisebox{0.0pt}{\makebox(0,0){\footnotesize $\circ$}}}
\put(647,404){\raisebox{0.0pt}{\makebox(0,0){\footnotesize $\circ$}}}
\put(657,404){\raisebox{0.0pt}{\makebox(0,0){\footnotesize $\circ$}}}
\put(668,422){\raisebox{0.0pt}{\makebox(0,0){\footnotesize $\circ$}}}
\put(678,422){\raisebox{0.0pt}{\makebox(0,0){\footnotesize $\circ$}}}
\put(689,481){\raisebox{0.0pt}{\makebox(0,0){\footnotesize $\circ$}}}
\put(699,554){\raisebox{0.0pt}{\makebox(0,0){\footnotesize $\circ$}}}
\put(710,553){\raisebox{0.0pt}{\makebox(0,0){\footnotesize $\circ$}}}
\put(720,553){\raisebox{0.0pt}{\makebox(0,0){\footnotesize $\circ$}}}
\put(731,556){\raisebox{0.0pt}{\makebox(0,0){\footnotesize $\circ$}}}
\put(741,541){\raisebox{0.0pt}{\makebox(0,0){\footnotesize $\circ$}}}
\put(752,565){\raisebox{0.0pt}{\makebox(0,0){\footnotesize $\circ$}}}
\put(762,565){\raisebox{0.0pt}{\makebox(0,0){\footnotesize $\circ$}}}
\put(773,614){\raisebox{0.0pt}{\makebox(0,0){\footnotesize $\circ$}}}
\put(783,600){\raisebox{0.0pt}{\makebox(0,0){\footnotesize $\circ$}}}
\put(794,586){\raisebox{0.0pt}{\makebox(0,0){\footnotesize $\circ$}}}
\put(804,578){\raisebox{0.0pt}{\makebox(0,0){\footnotesize $\circ$}}}
\put(815,728){\raisebox{0.0pt}{\makebox(0,0){\footnotesize $\circ$}}}
\put(825,766){\raisebox{0.0pt}{\makebox(0,0){\footnotesize $\circ$}}}
\put(836,661){\raisebox{0.0pt}{\makebox(0,0){\footnotesize $\circ$}}}
\put(846,663){\raisebox{0.0pt}{\makebox(0,0){\footnotesize $\circ$}}}
\put(857,662){\raisebox{0.0pt}{\makebox(0,0){\footnotesize $\circ$}}}
\put(867,567){\raisebox{0.0pt}{\makebox(0,0){\footnotesize $\circ$}}}
\put(878,596){\raisebox{0.0pt}{\makebox(0,0){\footnotesize $\circ$}}}
\put(888,615){\raisebox{0.0pt}{\makebox(0,0){\footnotesize $\circ$}}}
\put(899,567){\raisebox{0.0pt}{\makebox(0,0){\footnotesize $\circ$}}}
\put(909,526){\raisebox{0.0pt}{\makebox(0,0){\footnotesize $\circ$}}}
\put(920,523){\raisebox{0.0pt}{\makebox(0,0){\footnotesize $\circ$}}}
\put(930,521){\raisebox{0.0pt}{\makebox(0,0){\footnotesize $\circ$}}}
\put(941,555){\raisebox{0.0pt}{\makebox(0,0){\footnotesize $\circ$}}}
\end{picture}

%% file: OP_1-48.eepic
\setlength{\unitlength}{0.1400pt}
\begin{picture}(1500,900)(0,0)
\thicklines \path(174,67)(194,67)
\thicklines \path(1436,67)(1416,67)
\put(152,67){\makebox(0,0)[r]{-2}}
\thicklines \path(174,166)(194,166)
\thicklines \path(1436,166)(1416,166)
\put(152,166){\makebox(0,0)[r]{-1.5}}
\thicklines \path(174,264)(194,264)
\thicklines \path(1436,264)(1416,264)
\put(152,264){\makebox(0,0)[r]{-1}}
\thicklines \path(174,363)(194,363)
\thicklines \path(1436,363)(1416,363)
\put(152,363){\makebox(0,0)[r]{-0.5}}
\thicklines \path(174,461)(194,461)
\thicklines \path(1436,461)(1416,461)
\put(152,461){\makebox(0,0)[r]{0}}
\thicklines \path(174,560)(194,560)
\thicklines \path(1436,560)(1416,560)
\put(152,560){\makebox(0,0)[r]{0.5}}
\thicklines \path(174,658)(194,658)
\thicklines \path(1436,658)(1416,658)
\put(152,658){\makebox(0,0)[r]{1}}
\thicklines \path(174,757)(194,757)
\thicklines \path(1436,757)(1416,757)
\put(152,757){\makebox(0,0)[r]{1.5}}
\thicklines \path(174,855)(194,855)
\thicklines \path(1436,855)(1416,855)
\put(152,855){\makebox(0,0)[r]{2}}
\thicklines \path(174,67)(174,87)
\thicklines \path(174,855)(174,835)
\put(174,22){\makebox(0,0){0}}
\thicklines \path(437,67)(437,87)
\thicklines \path(437,855)(437,835)
\put(437,22){\makebox(0,0){10}}
\thicklines \path(700,67)(700,87)
\thicklines \path(700,855)(700,835)
\put(700,22){\makebox(0,0){20}}
\thicklines \path(963,67)(963,87)
\thicklines \path(963,855)(963,835)
\put(963,22){\makebox(0,0){30}}
\thicklines \path(1226,67)(1226,87)
\thicklines \path(1226,855)(1226,835)
\put(1226,22){\makebox(0,0){40}}
\thicklines \path(174,67)(1436,67)(1436,855)(174,855)(174,67)
\put(290,765){\makebox(0,0)[l]{1}}
\put(560,150){\makebox(0,0)[l]{$\gamma_0 D_0$}}
\put(660,350){\makebox(0,0)[l]{$\gamma_i D_0$}}
\put(980,615){\makebox(0,0)[l]{$\gamma_0 D_1$}}
\put(1068,530){\makebox(0,0)[l]{$\gamma_1 D_1$}}
\put(1042,320){\makebox(0,0)[l]{$\gamma_i D_1$}}
\put(200,461){\raisebox{-0.3pt}{\makebox(0,0){$\circ$}}}
\put(227,461){\raisebox{-0.3pt}{\makebox(0,0){$\circ$}}}
\put(253,461){\raisebox{-0.3pt}{\makebox(0,0){$\circ$}}}
\put(279,461){\raisebox{-0.3pt}{\makebox(0,0){$\circ$}}}
\put(305,660){\raisebox{-0.3pt}{\makebox(0,0){$\circ$}}}
\put(332,461){\raisebox{-0.3pt}{\makebox(0,0){$\circ$}}}
\put(358,461){\raisebox{-0.3pt}{\makebox(0,0){$\circ$}}}
\put(384,461){\raisebox{-0.3pt}{\makebox(0,0){$\circ$}}}
\put(411,461){\raisebox{-0.3pt}{\makebox(0,0){$\circ$}}}
\put(437,461){\raisebox{-0.3pt}{\makebox(0,0){$\circ$}}}
\put(463,461){\raisebox{-0.3pt}{\makebox(0,0){$\circ$}}}
\put(490,461){\raisebox{-0.3pt}{\makebox(0,0){$\circ$}}}
\put(516,461){\raisebox{-0.3pt}{\makebox(0,0){$\circ$}}}
\put(542,461){\raisebox{-0.3pt}{\makebox(0,0){$\circ$}}}
\put(568,461){\raisebox{-0.3pt}{\makebox(0,0){$\circ$}}}
\put(595,461){\raisebox{-0.3pt}{\makebox(0,0){$\circ$}}}
\put(621,238){\raisebox{-0.3pt}{\makebox(0,0){$\circ$}}}
\put(647,390){\raisebox{-0.3pt}{\makebox(0,0){$\circ$}}}
\put(674,390){\raisebox{-0.3pt}{\makebox(0,0){$\circ$}}}
\put(700,390){\raisebox{-0.3pt}{\makebox(0,0){$\circ$}}}
\put(726,461){\raisebox{-0.3pt}{\makebox(0,0){$\circ$}}}
\put(752,461){\raisebox{-0.3pt}{\makebox(0,0){$\circ$}}}
\put(779,461){\raisebox{-0.3pt}{\makebox(0,0){$\circ$}}}
\put(805,461){\raisebox{-0.3pt}{\makebox(0,0){$\circ$}}}
\put(831,461){\raisebox{-0.3pt}{\makebox(0,0){$\circ$}}}
\put(858,461){\raisebox{-0.3pt}{\makebox(0,0){$\circ$}}}
\put(884,461){\raisebox{-0.3pt}{\makebox(0,0){$\circ$}}}
\put(910,461){\raisebox{-0.3pt}{\makebox(0,0){$\circ$}}}
\put(936,461){\raisebox{-0.3pt}{\makebox(0,0){$\circ$}}}
\put(963,461){\raisebox{-0.3pt}{\makebox(0,0){$\circ$}}}
\put(989,461){\raisebox{-0.3pt}{\makebox(0,0){$\circ$}}}
\put(1015,461){\raisebox{-0.3pt}{\makebox(0,0){$\circ$}}}
\put(1042,557){\raisebox{-0.3pt}{\makebox(0,0){$\circ$}}}
\put(1068,477){\raisebox{-0.3pt}{\makebox(0,0){$\circ$}}}
\put(1094,389){\raisebox{-0.3pt}{\makebox(0,0){$\circ$}}}
\put(1121,389){\raisebox{-0.3pt}{\makebox(0,0){$\circ$}}}
\put(1147,461){\raisebox{-0.3pt}{\makebox(0,0){$\circ$}}}
\put(1173,461){\raisebox{-0.3pt}{\makebox(0,0){$\circ$}}}
\put(1199,461){\raisebox{-0.3pt}{\makebox(0,0){$\circ$}}}
\put(1226,461){\raisebox{-0.3pt}{\makebox(0,0){$\circ$}}}
\put(1252,461){\raisebox{-0.3pt}{\makebox(0,0){$\circ$}}}
\put(1278,461){\raisebox{-0.3pt}{\makebox(0,0){$\circ$}}}
\put(1305,461){\raisebox{-0.3pt}{\makebox(0,0){$\circ$}}}
\put(1331,461){\raisebox{-0.3pt}{\makebox(0,0){$\circ$}}}
\put(1357,461){\raisebox{-0.3pt}{\makebox(0,0){$\circ$}}}
\put(1383,461){\raisebox{-0.3pt}{\makebox(0,0){$\circ$}}}
\put(1410,461){\raisebox{-0.3pt}{\makebox(0,0){$\circ$}}}
\put(1436,461){\raisebox{-0.3pt}{\makebox(0,0){$\circ$}}}
\thinlines \path(200,461)(200,461)(227,461)(253,461)(279,461)(305,701)(332,461)(358,461)(384,461)(411,461)(437,461)(463,461)(490,461)(516,461)(542,461)(568,461)(595,461)(621,186)(647,401)(674,401)(700,401)(726,461)(752,461)(779,461)(805,461)(831,461)(858,461)(884,461)(910,461)(936,461)(963,461)(989,461)(1015,461)(1042,524)(1068,478)(1094,401)(1121,401)(1147,461)(1173,461)(1199,461)(1226,461)(1252,461)(1278,461)(1305,461)(1331,461)(1357,461)(1383,461)(1410,461)(1436,461)(1436,461)
\end{picture}